# ASSESSMENT OF THE AREA MEASUREMENT ON CARTOSAT-1 IMAGE

Joanna Pluto-Kossakowska, David Grandgirard, Rafał Zieliński, Simon Kay

Joint Research Centre, Institute of Protection and Security of Citizen

Via Fermi 1, 21027 Ispra, Italy

Commission I, SS-11

KEY WORDS: accuracy, agriculture, buffer tolerance, Cartosat-1, measurement, parcel area, photointerpretation

ABSTRACT:

The goal of this study was the evaluation of agriculture parcel area measurement accuracy on Cartosat-1 imagery, and the determination of the technical tolerance appropriate for measurement using photointerpretation techniques. A further objective was to find out the influence of image type, land cover or parcel size on the area measurement variability. In our experiment, five independent operators measured 185 parcels, 3 times, on each image. Next, the buffer width, calculated as the difference between measured and reference parcel area, was derived and was the subject of statistical analysis. Prior to verifying the normality of the buffer widths, a detection of anomalous measurements is recommended. This detection of outliers within each group of observations (i.e. parcels) was made using the Jacknife distance test on each type of imagery (Cartosat Aft, Cartosat Fore). Then, the General Linear Model procedure to identify major significant effects and interactions was followed by analysis of variance to ease the interpretation of the variability observed of the area measurement. Finally, two different parameters, reproducibility limit and critical difference, were calculated to make comparison with other sensors like digital aerial orthophoto in this study possible. The repeatability limits gave the acceptability difference between two operators when measuring the same parcel. For orthophoto this value reached 2.86m, on Cartosat-1 5.17m and 8.76m for Aft and Fore image respectively.

## 1. INTRODUCTION

In the frame of the European Common Agricultural Policy (CAP), a large number of agricultural parcels claimed for subsidies have to be measured with high accuracy every year by the Member States administrations. This is usually done using Very High Resolution (VHR) optical images with ground sampling distance (GSD) of around 1m or better. In 2007, about 50% of these area measurements were carried out using VHR optical satellite imagery acquired over 150 000 km2 and 220 zones. To account for the uncertainty inherent to any measurement tool, a technical tolerance is used when comparing the claimed and measured areas to determine the validity of the claimed area. According to article 30 of EC Reg. N° 796/04, this technical tolerance must not exceed a buffer of 1.5 m applied to the perimeter of the agricultural parcel. Due to this specification set on the tolerance, orthorectified VHR images with resolution less than or equal to 1m are normally to be used for agricultural parcel measurement. Since the acquisition of precise imagery may fail because of weather conditions or competition between clients, the images with ground sampling distance around 2m are also used, but generally as back-up. It is therefore worth examining the potential of such data (Cartosat-1) for replacing other VHR images in the context of European Commission requirements.

The purpose of the research is to discover the extent to which the area from independent measures can vary in practice and why. Therefore, the scope of this study is twofold: first, to assess the agricultural parcels identification capabilities from Cartosat-1 against digital orthophoto and second, to evaluate the area measurement accuracy and buffer tolerance for Cartosat-1 image when using Computer Assisted Photointerpretation (CAPI) techniques. To supplement the evaluation of measurement on Cartosat-1 image, the analysis of influence of different factors on buffer variability was carried out.

2. MATERIAL & METHODS

2.1 Data

Following the good results from the previous study on geometric precision of Cartosat-1 in the frame of Cartosat-SAP Working Group (Kay, Zieliński 2006, Crespi et al. 2006), the parcel area measurement was proposed on orthorectified stereo pairs from 31st of January 2006 for the Mausanne site. Cartosat-1 carries two panchromatic (0.50-0.85 µm) cameras which acquire stereoscopic imagery. The cameras make possible nearly simultaneous imaging of the same area from two different angles: Fore with a tilt of +26deg and Aft with a tilt of -5deg from the yaw axis. Instantaneous geometric field is 2.5m at nadir. The orthorectification process was carried out using the Leica Photogrammetic Suit (LPS) software. The RPC approach was applied to Cartosat-1 stereo pair, using six well established, regularly distributed ground control points and a highly accurate reference DEM. This DEM presents a verified quality (linear Root Mean Square Error [RMSE] in the vertical axis, Z) of better than 0.60m on well defined points. The data have a ground sample distance (GSD) of 2m. The final planimetric accuracy of orthorectified images (Fore and Aft) was checked against 25 independent check points and the linear RMSE of around 1 pixel (that is, either X or Y directions) was reached (Kay, 2006).

In order to assess the area measurement accuracy, reference parcels with a known area and perimeter have to be selected from available sources or acquired using independent tools, here precise orthophoto. This aerial orthophoto with 0.5m of GSD was acquired in 14th of May 2005 as multispectral image RGB using UltraCamD digital camera. Pixel level accuracy was determined in a separate experiment (Spruyt, pers. comm.).

2.2 Methods

The methodology of area measurement evaluation is based on statistical analysis of discrepancies between the measured and reference areas (Pluto-Kossakowska et al. 2007). In order to derive the tolerance above which an inspector will reject the area claimed by the farmer with a risk of α=5%, an initial verification that the distribution of the buffer is normal must be made. To obtain the final tolerance for the measurements, a repeatability limit of the buffer was applied.

The scheme of validation procedure is proposed as followed:

    1. Data processing and acquiring:

        - Images orthorectification – needed to make cartographic product to measure

        - Acquiring the reference parcels – from digital orthophoto

        - Area measurement of the selected parcels on the images

        - Buffer calculation based on measurements and reference data

    2. Statistical analysis of the buffer value

- Anomalous measurements detection and elimination

- Normality test and analysis of variance (SLS, ANOVA)

- Determination of tolerance for the measurements as reproducibility limits.

It is practical to model the maximum acceptable discrepancy between the measured area and the claimed area, i.e. the tolerance, as the parcel perimeter multiplied by a width. This width, also called buffer width (or simply buffer) around the parcel perimeter, is expected to vary as a function of the measurement tool, whether it is an image or a GPS-device. For a given parcel, the knowledge of its reference (i.e. true) area and reference perimeter allow the transformation of the area error (measured area – reference area) into this buffer width using:

$$B_i = \frac{(a_i - a_{ref})}{p_{ref}}$$ (eq. 1)

where  $B_i$ = buffer width for measurement i
$a_i$ = measured area for measurement i
$a_{ref}$ = reference area of the parcel
$p_{ref}$ = reference perimeter of the parcel

Using the buffer values from different observations we can determine the tolerance between two independent measurements under the specified condition (the same parcels, same image, and independent operators). The simplest way is to verify whether the distribution of the buffers follows a normal law using different tests.

Detection of outliers is recommended prior to verifying the normality of the buffer widths. According to ISO 5725 (1994), the detection of anomalous measurements may be made using different tests. Outlier detection was performed here using the Jacknife distance test in JMP 6.0 (SAS Institute). A Standard Least Square (SLS) procedure was then performed to identify the factors (and 2nd/3rd order interactions) significantly explaining the observed variability of the buffer. Table 1 presents the list of factors and related modalities. "Shape" factor was distinguished on to three modalities: simple i.e. rectangular alike shape, medium i.e. rectangular shape with little changes and complex i.e. shapeless. The visibility depends on parcel itself, parcel surrounding and image properties: good visibility i.e. all parcel borders are easy to recognise; poor visibility – part of the border is difficult to recognise and must be deduced. "Operator" presented two different modalities (skilled vs. unskilled) based on the level of "experience" of each photointerpreter had at the beginning of the survey.

Finally, the assumption of normal distribution of the buffer values leads to the derivation of a tolerance (at α=5%), above which an inspector would reject the area claimed by the farmer. For the needs of our survey and following the ISO 5725 (1994), the tolerance can be interpreted as reproducibility limit (eq. 2). Reproducibility refers to the ability of the measurement to be accurately reproduced by someone else working independently, i.e. is a value less than or equal to which the absolute difference between two results obtained under reproducibility conditions may be expected to be with a probability of x%.

$$R = f * \theta_R * \sqrt{n}$$ (eq. 2)

Where  $\theta_R$ = standard deviation under reproducibility conditions (for the method of calculation refer to ISO 5725, 1994)

$f$ = multiplication factor of standard deviation to determine the confidence interval on specified level of probability (here 95%)
$n$ = number of test results to be compared, here n=2

For normal distribution at 95% probability level, f is 1.96 and f*√2 then is 2.77. The simple "rule of thumb" R=2.8σ$_R$ is applied instead of equation (2) (ISO 5725, 1994).

| Factors | Modalities |
|---|---|
| Operator (n=5) | Skilled (n=3) |
|  | Unskilled (n=2) |
| Image (n=3) | Orthophoto |
|  | Cartosat-1 Aft |
|  | Cartosat-1 Fore |
| Image visibility (n=4) | Good on all images |
|  | Good on ortho, poor on cartosat |
|  | Poor on ortho good on cartosat |
|  | Poor on all images |
| Parcel shape (n=3) | Simple |
|  | Medium |
|  | Complex |
| Parcel size (n=3) | Small (< 2ha) |
|  | Medium (2ha> >8ha) |
|  | Large (>8ha) |
| Land cover type (n=7) | Bare soil |
|  | Green cover |
|  | Marsh |
|  | Olive trees |
|  | Orchard |
|  | Pasture |
|  | Vineyard |

Table 1. List of factors tested and related modalities

The other parameter which was used to evaluate the test results is critical difference (CD) to the reference value. This value is a value less than or equal to which the absolute difference between particular observation and reference data is expected to be at the 95% of confidence (ISO 5725). The critical difference as comparison with reference value for more than one operator can be calculated as below (3):

$$CD = \frac{1}{\sqrt{2p}} \sqrt{(2.8\theta_R)^2 - (2.8\theta_r)^2(1 - \frac{1}{p}\sum\frac{1}{n_i})} \qquad \text{(eq. 3)}$$

Where  $\theta_R$ = standard deviation under reproducibility conditions
$n$ = no of observations
$p$ = no of operators
$\theta_r$ = standard deviation under repeatability conditions

3. MEASUREMENT

The test site was situated on the region of Maussane in France and covered 10km by 10km area. All data, i.e. stereo pair of Cartosat-1 Aft and Fore and digital orthophoto UltraCamD, were projected in UTM 31N (WGS 1984). The set of parcels (as polygons) was acquired using ARCGIS software as the measuring environment.

The number of parcels to be measured was probabilistically determined according to previous results (Pluto-Kossakowska et al., 2007). After field verification (Feb 2008), 203 parcels were digitised on UltracamD RGB composition, then verified and corrected by independent operator. This set of parcels was checked also on Cartosat-1 images, and 18 parcels were rejected in terms of invisible borders. This means that only 8% of 203 parcels were not detectable on Cartosat-1. Finally, 185 parcels were used for the study. The land cover classes and parcel borders were precisely identified on these selected parcels during the field campaign in February 2008.

Three replicates per parcel were obtained from five operators on two images of Cartosat-1 (Aft and Fore) and on the UltraCamD orthophoto. The parcels appeared on the screen randomly and were enlarged at the maximum zoom within the viewer window. On screen, the parcel to be measured was centrally identified (see figure 1). After the parcels measurement, the area and perimeter were calculated for each observation. The buffer value was derived using equation (1) and was used for following statistical analysis.

4. STATISTICAL ANALYSIS AND RESULTS

4.1 Outliers

The workflow included a statistical detection of outliers, a great variety of which are available from the literature and which could be used in this experiment. Using the Jacknife distances test, 81 observations out of 4995 (1.6%) were identified as significant. SLS procedures allowed the identification of the factors and related interactions responsible of the outliers' population distinction (($F(6;80)=30.51$ p-value<0.0001, $r^2=0.71$). The maximum value of the outliers was 61,2m while the minimum value was    -79,2m.

The majority of the outlying observations occurred once per parcel. Less frequent were the parcels several times with multiple outliers: these concerned essentially the three replicates from the same operator (and the same image). Even if existing, it was very rare that the same image was identified as an outlier by several operators from several images. To illustrate this specific case, an example is given in Figure 1.

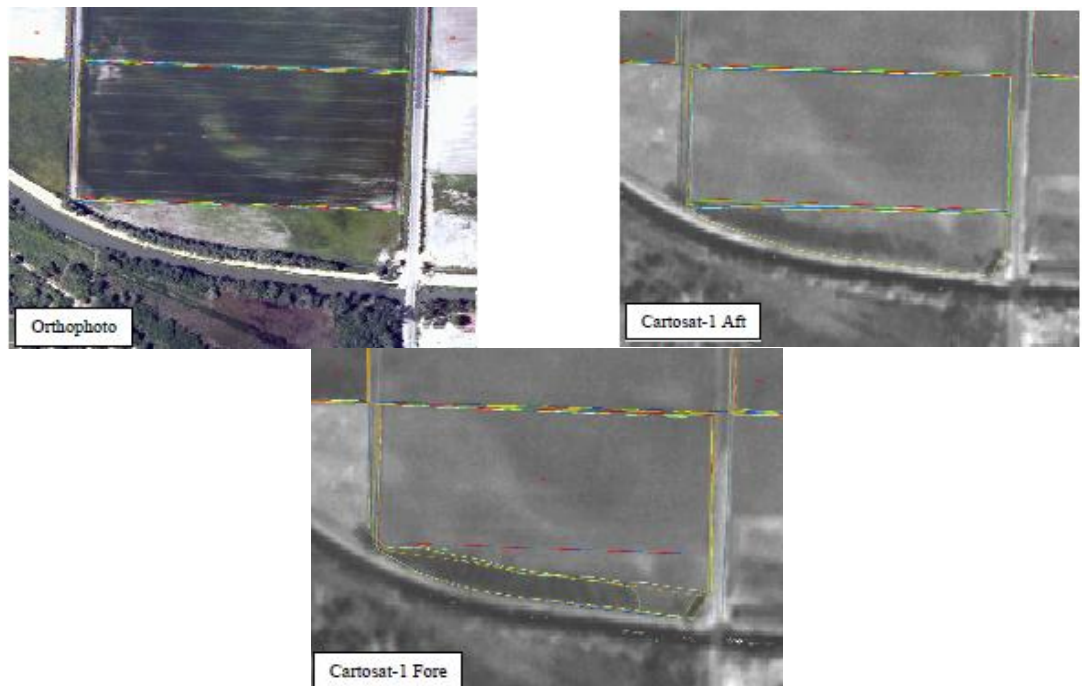

Figure 1. Example of outliers: 1 on Cartosat Aft, 6 on Cartosat Fore, none on orthophoto

The two main groups of factors explaining outliers were (i) the "parcel # image properties" defined as the visual representation of parcel characteristics within an image and (ii) the level of experience of one operator when interpreting image.

Concerning "parcel # image properties", only "image" (F=6.11 p<0.0157) and "parcel size" (F=58.06 p<0.0001) were single factors significantly affecting the area buffer. Then, 2nd order interactions between the shape of the parcel ("image* parcel shape" F=33.79 p<0.0001), the size of the parcel ('image * parcel size" F=11.83 p=0.001) and "image" were significant. From ANOVA, orthophoto counted for 23% of the outliers and was significantly underestimated (mean value = -16.7m; F=6.50 p=0.0025) when Cartosat-1 Aft and Fore counted respectively for 18% and 58% with mean values of -2.6m and +7.5m.

As discussed in the next paragraph, this result did not signify that the orthophoto is a major source of underestimation (or Cartosat-1 Fore as a source of overestimation). In fact, the operator's interpretation was mainly the source of underestimation. Concerning parcel size, small and large parcels counted respectively for 22% and 26% of the outliers when medium parcels represented 56% of the outliers. Difference of buffer was significant between parcel sizes (F=26.2 p<0.0001) with mean buffer values of +12.6m, +8.5 and -28.1m respectively for small, medium and large parcels. Independently of the image, of the parcel shape or of the operator, small parcel areas were often overestimated and large parcel areas were underestimated. One would suggest that this effect was a consequence of the magnification function used during the measurement: each parcel was enlarged and fitted to the full screen size to facilitate boundary recognition. The smaller the parcels, the higher the magnification was and the higher the dilution of the contrast. On the other hand, one could imagine a "compensation" effect from the operator which, unconsciously, searched to not disadvantage small parcels and tended to put boundaries off and inversely.

As explained previously, few outliers came from orthophoto when most of them were due to panchromatic Cartosat-1 images: Cartosat-1 Aft and Fore images were respectively responsible of 18.5 and 58% of the outliers detected. Clarity of the object displayed on the screen and true colour (RGB) composition seemed essential first for a good recognition of the object, second for the better delineation of the parcel boundaries. Thus, "operator's recognition capacity" and "operator's object memorisation" could be of prime importance, at least to explain outlier's existence. They both compose "operator's experience" and condition the interpretation of the size and shape of the parcel (and of possible contained object to be excluded). With Cartosat-1 images, we assumed that parcel boundaries recognition was more difficult and more deductive for operator. We showed that "operator*image" effect on buffer measurement was significant (F=3.6 p=0.018). Over the five operators, four were responsible for the outliers; among those, three were skilled. The unskilled operator was responsible of most of the outliers (70%) and generally underestimated the buffer (mean value = -6.2m) when others always overestimated the buffer (from +12.2 to +14.6m). All together, the results obtained from the analysis of the outliers' population clearly suggested a tripartite relationship existing between (i) the parcel with its particular characteristics, (ii) the image as the information vector conveying parcel characteristics and (iii) the operator as the place of interpretation where his personal visual recognition capacity and object memorisation interact and determine the accuracy of the measurement. Here we showed that the unskilled operator was mainly responsible of underestimation of parcel area certainly because of a too limited object recognition experience on image, this, whatever the type of image considered. On the contrary, the three skilled operators often overestimated parcel areas. They were responsible for the major part of overestimation on Cartosat-1 image. This should be related to the fact that their experience initially concerned true colour composition imagery. Consequently, (i) effect of magnification, (ii) unconscious parcel size related "compensation effect" and/or (iii) lost of reference when passing from orthophoto to panchromatic could explain outliers from skilled operators.

By analysing extreme area discrepancies, we showed that area measurement accuracy is mainly conditioned by the relationship between operator and the image properties. CAPI training for panchromatic and true colour images should be organised to reduce risk of wrong delineation of the agricultural parcels.

4.2 Final buffer population

4.2.1 Normality test

A normality test (KSL/Kolmogorov-Smirnov-Lilliefors) was performed on each type of image to determine if observed buffer values were normally distributed and consequently to decide if an analysis of variance was relevant to identify the main significant factors and differences between modalities. Whatever the type of image, small p-values were obtained (W=0.951 p-value<0.001; W=0.989 p-value<0.001; W=0.996 p-value=0.001 for Orthophoto, Cartosat1 Fore and Aft respectively) and the null hypothesis that observed buffer values have a normal distribution were rejected. However, buffer value distribution was very close to the normal distribution, being relatively symmetrical around the mean. Furthermore, the number of observations by type of image was sufficiently high (Table 2) to allow for processing SLS procedures and analysis of variance without restriction and to limit the risk of misinterpretation of factors' effects.

Buffer values obtained from the survey were significantly different between image modalities; mean values and standard deviation are given in table 2. Consequently, the SLS procedure and analysis of variance have been conducted for each type of image independently and will be discussed separately.

| Parameters [m] | Orthophoto | Cartosat Aft | Cartosat Fore |
|---|---|---|---|
| Mean | -0.059 | 0.041 | 0.515 |
| Std Dev | 0.990 | 1.848 | 2.857 |
| Std Err Mean | 0.024 | 0.046 | 0.074 |
| Upper 95% Mean | -0.011 | 0.131 | 0.661 |
| Lower 95% Mean | -0.106 | -0.049 | 0.370 |
| Number of observations | 1642 | 1613 | 1485 |

Table 2. Distribution parameters for each image

4.2.2 SLS and ANOVA: main significant effects

The SLS procedure applied separately to each image dataset showed that the factors responsible for the observed variability of the buffer were different between images. Whereas "operator", "shape" and "size" of the parcel or even "visibility" were the single factors to be retained for the orthophoto, "land cover" should also be considered of interest for Cartosat-1 Fore. Concerning Cartosat-1 Aft, only "visibility" was retained as the single factor influencing the buffer. All together, only "visibility" was significant and common to the three image datasets (F=3.875 p=0.009). Mean buffer value was generally significantly underestimated (-0.33m ± 1.75SD) for the third visibility modality "Poor on Ortho, good on Cartosat-1" whereas it was generally overestimated for the three other modalities (global mean of the three remaining modalities = 0.17m ± 2.04SD).

With regard to the orthophoto and Cartosat-1 Fore images, significant factors and interactions were very similar. Only "land cover" had a significant effect (F=24.29 p=0.0064) on buffer for Cartosat-1 Fore. For these two images, variance of the buffer was high between operators. Unskilled operators generally delineated the parcel boundary well on the orthophoto, but largely underestimated the parcel area on Cartosat-1 Fore (-0.05m ±0.99SD and -1.92m ±2.26SD respectively). Conversely, skilled operators tended to overestimate the buffer with Cartosat-1 (+2.56m ±2.13SD) but to generally delineate parcels correctly from the orthophoto. Once again, "magnification effect", "unconscious compensation effect" and/or loss of reference when passing from orthophoto to panchromatic could explain overestimation of parcel area by skilled operators.

Generally the larger the parcel, the smaller the overestimation of the parcel area by skilled operators on Cartosat-1 Fore. On the orthophoto, the smaller the parcels, the higher the difference between the digitised area and the true parcel area. On the contrary, parcel size didn't really influence the measurements for unskilled operators (F=0.93 p=0.39): parcel area was consistently underestimated when using Cartosat-1 Fore and relatively well measured with orthophoto, irrespective of the parcel size. Regarding the shape of the parcels, only a limited effect on the measurement of the parcel area (and subsequently of the buffer value) was observed. Parcel shape had a greater influence on buffer measurement when interacting with "image" and "parcel size". This was especially true for skilled operators for whom a complex parcel shape led to significant overestimation of the parcel area and consequently to higher positive buffer values. With regards to Cartosat-1 Aft, for which "visibility" was the only significant factor, numerous 2nd order interactions were significant. These principally concerned the shape and the size of the parcels, then the operators and finally the land cover.

Firstly, from the previous results concerning image types, we showed that the characteristics of the parcel clearly influenced the precision of the parcel area delineation: shape and size of a parcel, either separately or combined, are interpreted differently depending on the operator's experience. For experienced operators, large and/or complex parcels boundaries are generally smoothed because of the magnification effect or possibly as a consequence of productivity criteria (i.e. to cost-effectively process a maximum quantity of parcels a day), thus leading to overestimation of the parcel area. On the contrary, unskilled operators seemed to be less influenced by the parcel characteristics and constantly underestimated the parcel area. Secondly, image quality, as defined by the "visibility" factor, strongly influenced the final accuracy. Skilled operators used to working with orthophoto obtained relatively good results with orthophoto, but they tended to lose this advantage when switching to panchromatic images. Conversely, unskilled operators were frequently inaccurate with both orthophoto and with panchromatic images. Consequently, the use of one type of image cannot be decided without knowing the staff competences by assessing their abilities to transfer and use memorised CAPI experience. Therefore a first question could concern the best way to choose an operator according to the image type. When CAPI has to be performed on orthophoto or panchromatic images, a cost-effective solution would be to choose skilled operators, but with the risk that smoothing (complex/large) and compensation (complex/small) effects will generally lead to overestimation of the true area. On the other hand, if the strategy of the enterprise is to contract new photointerpreters, we suggest that one should assess their recognition capacity; this could be undertaken on true colour composite images regularly compared to their panchromatic equivalent.

| Image | Carto Aft | Carto Fore | Orthophoto |
|---|---|---|---|
| Mean Value = bias [m] | 0.04 | 0.52 | -0.06 |
| St. Dev. Repeatability [m] | 1.85 | 2.22 | 0.92 |
| Repeatability Limit [m] | 5.18 | 6.23 | 2.59 |
| St. Dev. Reproducibility [m] | 1.85 | 3.13 | 1.02 |
| **Reproducibility Limit [m]** | **5.17** | **8.76** | **2.86** |
| **Critical difference to reference [m]** | **1.65** | **3.21** | **0.96** |

Table 4. Results from area measurement on orthophoto and Cartosat-1

Intentionally, a last factor has not yet been discussed: land cover. The decision was made to discuss it separately so as not to risk overloading results or incorrectly classifying the main factors to consider from this survey. Indeed, land cover appeared to be significant only within 2nd order interactions and never as a single significant factor, suggesting that land cover cannot be discussed independently of other factors. From the SLS results, we showed that land cover was mainly associated with "visibility", "parcel size" and "parcel shape"; this indicated that land cover could be perceived as a characteristic of the object, i.e. the parcel, at the same level as "shape" and "size". Whatever the operator and his level of experience, we showed that parcel area measurement was always more accurate and less variable when there was bare soil, annual crops or pastures. On the contrary, for orchards, vineyards or olive trees, parcel area was often overestimated and highly variable. This was true especially with Cartosat-1 images. For instance mean buffer values were 0.44m ±1.27SD, 0.50m ±2.67SD, 0.61m ±3.64SD for orchards and -0.13m ±0.68SD, -0.08m ±1.34SD, 0.54m ±0.15SD for bare soil, respectively with orthophoto, Cartosat-1 Fore and Cartosat-1 Aft. This trend was maintained between operators, the sole difference being that unskilled operators continued to proportionally underestimate parcel area according to image type. When considering parcel size, larger parcels with bare soils, annual crops or pastures were often overestimated than small parcels. On the other hand, ligneous crops appeared to be the main source of underestimation of the area of small parcels. Finally, regarding parcel shape, the same results were obtained: greater difference from the true area and greater variability was evident for parcels with ligneous crops.

From these results, land cover seemed to aid the operator in the correct identification of a parcel regarding its content; it allowed the operator to recognise more clearly the parcel but it remained relatively useless when delineating the parcel boundaries. Tree canopies extending outside of the parcel could lead to overestimation of the area because of the difficulty of clearly distinguishing the parcel area and surrounding natural vegetation; and crops boundaries were delineated more often than parcels boundaries. We called it the ligneous overestimation effect.

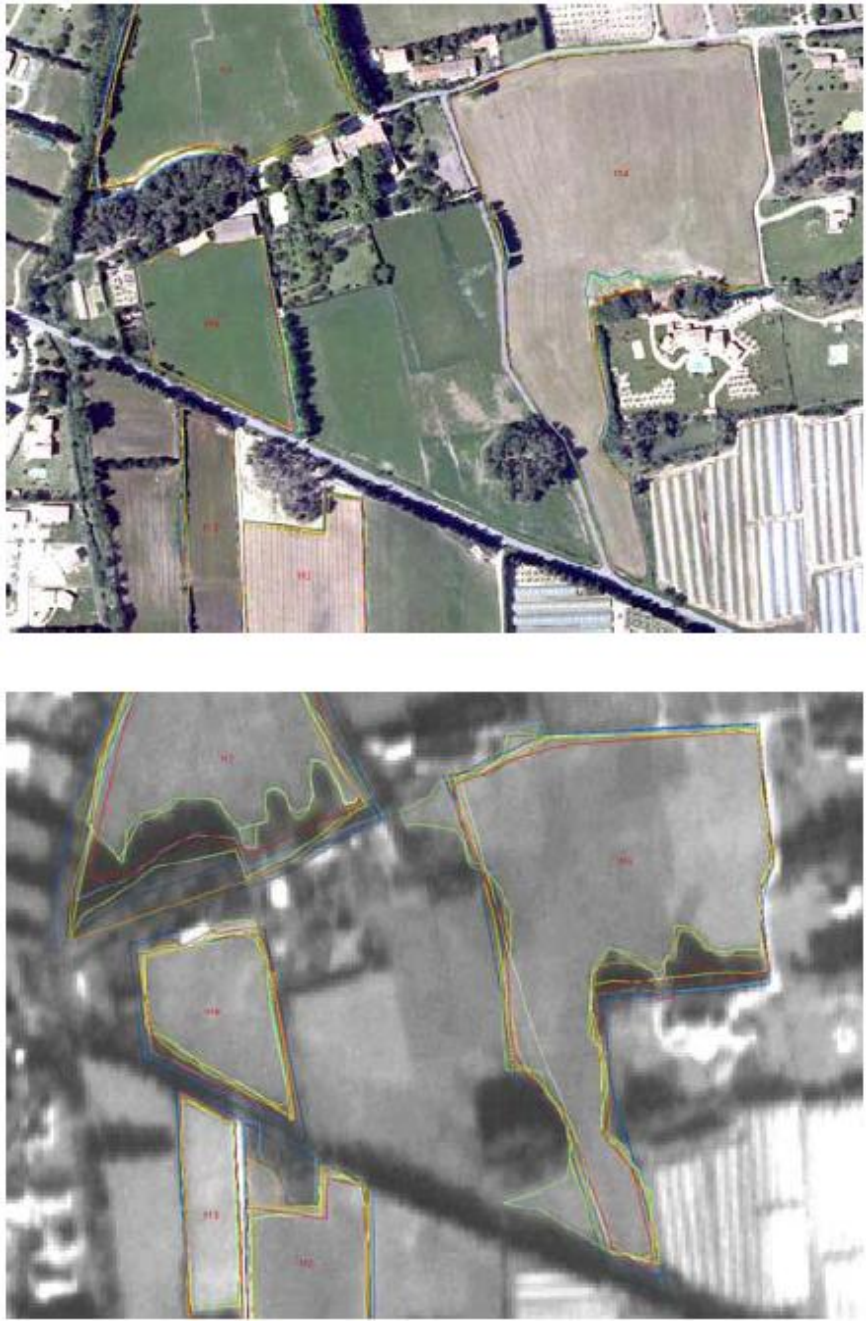

Figure 3. Example of area measurement on orthophoto and Cartosat1 images

4.2.3 Reproducibility and critical difference

The final results of buffer determination are presented in table 4. The repeatability limits gave the maximum difference between two operators on the same image at 95% confidence, considered as tolerance in this experiment. For orthophoto, this value was equal to 2.86m and was equal to 5.17m and 8.76m respectively for Cartosat-1 Aft and Fore. Even if Fore had higher value of tolerance than Aft, it didn't signify that Fore was worse than Aft. Effectively, Fore having been processed before Aft, we could assume that a "training" effect would have influenced the operators' recognition and memorisation capacities and limit the relevance of the comparison.

The critical difference (CD) value gave us the maximum difference between the reference area and the measured areas (again at 95% confidence). The lowest value was equal to 0.96m and was obtained from orthophoto. On Cartosat-1, CD was equal to 1.65m and 3.21m respectively with Aft and Fore. Regarding the two last results, Cartosat-1 images were both less accurate than orthophoto; and we consequently do not recommend using Cartosat-1 images as the main tool to perform CwRS under the actual CAP regulatory framework.

To illustrate the major limitations observed during this survey, figure 3 presents problems met with parcel border identification. On the orthophoto, long shadows representing vegetal hedges, which should not be considered as parcel boundaries, are visible; however, these shadows have been regularly considered as boundaries by operators in Cartosat-1 images and not included in the parcel area measurement. In addition, changes in texture for Cartosat-1 provoked the disappearance of narrow paths and resulted in overestimation.

## 5. FINAL DISCUSSION

Compared to the orthophoto, the majority of parcels were correctly identified on both Cartosat-1 images; only 62 observations out of 3330 were found to be outliers. Nevertheless, the main problem with parcel area measurement was the correct border identification due to a loss of information as a result of shadows, small and narrow objects, and texture changes. Overall, changes resulted in a less accurate delineation of the parcel boundaries and very often to an overestimation of the parcel area. With regard to reproducibility limits and critical difference, neither of the Cartosat-1 images tested can be considered as a primary solution for Control with Remote Sensing in accordance with European CAP requirements.

Comparison of images and evaluation of factors highlighted the need to consider the CAPI process as organised around a tripartite relationship between (i) global image quality, (ii) operator's recognition capacity and (iii) operator's object memorisation. As proposed in figure 4, these three components should be considered as integrated and dependent inside the CAPI system. Image interpretation and agricultural parcel boundary delineation appeared closely related to the operator's personal experience. CAPI experience means the capacity to recognise an object whatever the source of the information (the image) and the capacity to compare this information to a pool of personal reference obtained from regular training and/or previous experiences. Both initially depend on the global quality of the image; quality can be perceived as the effectiveness of the image to provide to the operator properties of each single object (e.g. precision) as well as difference between objects (e.g. contrast). The fact that "visibility" was one main problem when identifying parcel boundaries may confirm this assumption. Further, the operator's interpretation was often biased due to intrinsic parcel characteristics such as shape, size or land cover. This suggests that despite using images of high quality, CAPI efficiency remains dependent on the operator's references and his adaptability. To perform CAPI efficiently, any contractor responsible for CwRS should use a variety of the image sources when training its staff and regularly test the accuracy (deviation) of each individual involved within the process. By evaluating the relation between operator and images, land cover, physical characteristics of parcels, the contractor would efficiently assess the quality of the whole workflow, excel in measuring agricultural parcel area and quickly meet the diverse CwRS regulatory requirements.

Before that, several potential effects identified during this survey should be apprehended: the magnification or compensation effect, the smoothing effect and the ligneous over-estimation effect.

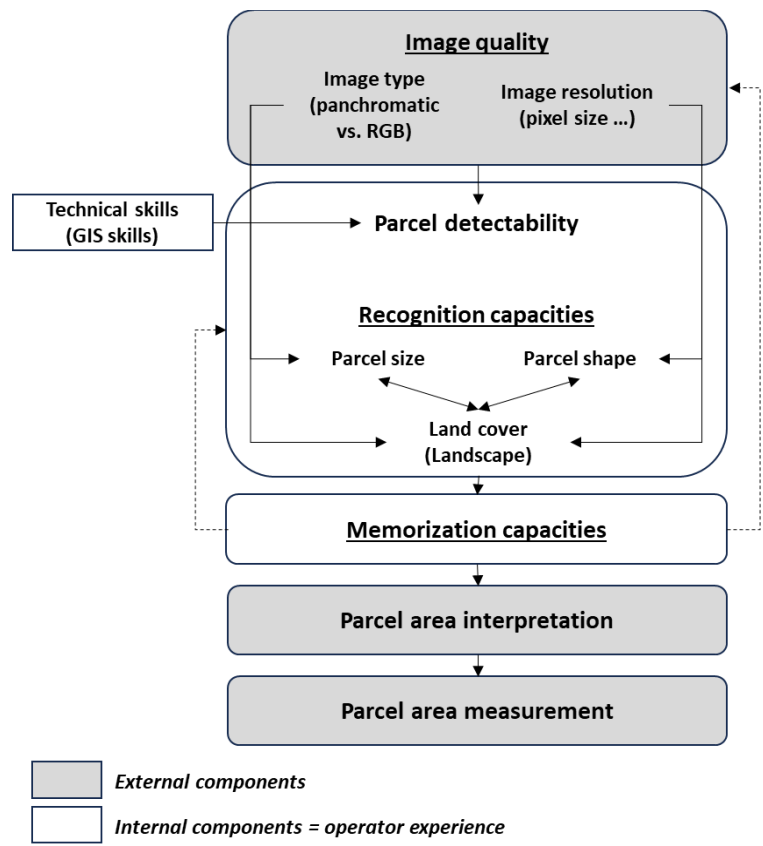

Figure 4: Organisation of the three CAPI components concerning agricultural parcel's area measurement